\documentclass{vgtc}                          

\graphicspath{{figures/}{pictures/}{images/}{./}} 

\usepackage{times}                     

\usepackage{tabu}                      
\usepackage{booktabs}                  
\usepackage{lipsum}                    
\usepackage{mwe}                       

\usepackage{mathptmx}                  
\usepackage{algorithm}
\usepackage{algorithmicx}
\usepackage{algpseudocode}
\usepackage{listings}
\usepackage{amsmath,amssymb,amsfonts}
\usepackage{subscript}
\usepackage{multirow} 

\onlineid{0}

\vgtccategory{Research}

\vgtcinsertpkg

\title{MExplore: an entity-based visual analytics approach for medical expertise acquisition}

\author{Xiao Pang\thanks{e-mail: xiaopang@scu.edu.cn}\\ %
        \scriptsize  West China Hospital of Stomatology\\ Sichuan University %
\and Yan Huang\thanks{e-mail: 464311079@qq.com}\\ %
     \scriptsize West China Hospital of Stomatology\\ Sichuan University %
\and Chang Liu\thanks{e-mail: liu$\_$chang$\_$92@sina.com}\\ %
     \scriptsize West China Hospital of Stomatology\\ Sichuan University %
\and JiYuan Liu\thanks{e-mail: wchsljy@scu.edu.cn\textsuperscript{*}}\\ %
     \parbox{1.4in}{\scriptsize \centering West China Hospital of Stomatology\\ Sichuan University}
\and MingYou Liu\thanks{e-mail: liumingyou@pumch.cn\textsuperscript{*}}\\ %
     \parbox{1.4in}{\scriptsize \centering Peking Union Medical College Hospital\\ Chinese Academy of Medical Sciences}
}

\teaser{
  \centering
  \includegraphics[width=\linewidth]{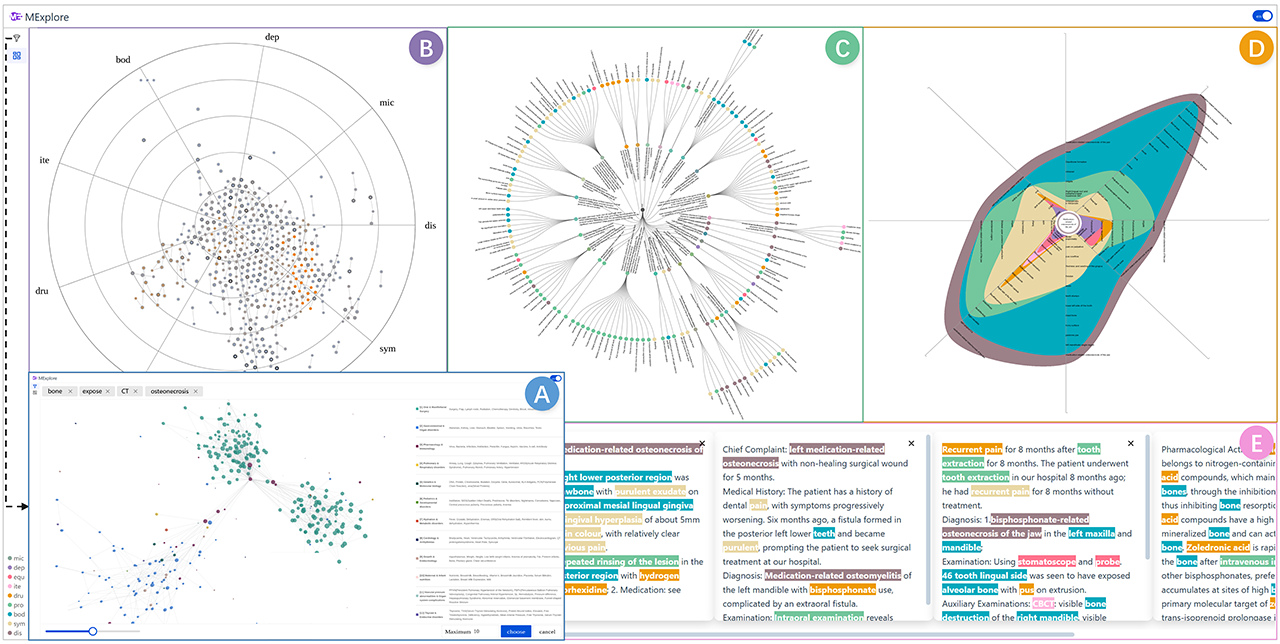}
  \caption{Overview of the MExplore system. \textbf{A} Displays the spatial and thematic distribution of the document, allowing users to select the exploration range. \textbf{B} Analyzes paragraph subgraphs to identify areas of interest for further exploration. \textbf{C} Visualizes the relationships between entities for a comprehensive understanding. \textbf{D} Focuses on a specific entity for detailed, in-depth analysis. \textbf{E} Provides access to the original text to ensure authenticity and support cyclical analysis.}
  \label{fig:mainInterface}
}

\abstract{
	Acquiring medical expertise is a critical component of medical education and professional development. While existing studies focus primarily on constructing medical knowledge bases or developing learning tools based on the structured, private healthcare data, they often lack methods for extracting expertise from unstructured medical texts. These texts constitute a significant portion of medical literature and offer greater flexibility and detail compared to structured data formats. Furthermore, many studies fail to provide explicit analytical and learning pathways in this context.
	This paper introduces MExplore, an interactive visual analytics system designed to support the acquisition of medical expertise. To address the challenges of the inconsistencies and confidentiality concerns inherent in unstructured medical texts, we propose a workflow that employs a fine-tuned BERT-based model to extract medical entities (MEs) from them. We then present a novel multilevel visual analysis framework that integrates multiple coordinated visualizations, enabling a progressive and interactive exploration of medical knowledge.
	To assess the effectiveness of MExplore, we conducted three case studies, a user study, and interviews with domain experts. The results indicate that the system significantly enhances the medical expertise acquisition process, providing an effective interactive approach for acquiring and retaining knowledge from medical texts.
} 


\begin{document}


\firstsection{Introduction}

\maketitle
Medical expertise is the highly organized and differentiated knowledge base that allows experts to see the world in different ways from non-experts. There is highly stratified expertise from novice medical students to board-certified medical specialists. As medical education increases, the number of medical students each year increases\cite{NHC2025, AAMC2023}. This trend reflects the growing demand and challenges for medical expertise acquisition.

The acquisition of medical expertise is a complex cognitive process that involves constructing abstract knowledge networks and iteratively refining them into narrative structures known as ``illness scripts''\cite{schmidt1993acquiring}. These scripts, which encapsulate diagnostic reasoning—cognitive processes through which clinicians gather, analyze, and synthesize information to identify the most likely diagnosis. Illness scripts operationalise this reasoning by organising experiential knowledge into memorable narratives that link enabling conditions, pathophysiological faults, and clinical consequences\cite{charlin2007scripts}.

Medical texts, which are usually organized as medical documents (MDs), such as electronic medical records (EMRs), serve as the primary repository for this case-based knowledge, offering rich, detailed accounts of real-world clinical events. However, the sheer volume and unstructured nature of these texts present a formidable challenge to knowledge extraction. While recently, Large Language Models (LLMs) have emerged as a powerful tool for knowledge acquisition, but their direct application in medicine is fraught with risk, where LLMs are prone to exhibit ``deceptive expertise'',  generating plausible yet factually incorrect information while failing to recognize their own limitations \cite{griot2025large, kim2025medical}. Over-reliance on these models can also foster cognitive offloading, leading to cognitive decline \cite{safranek2023role, dergaa2024tools}, impairing their capacity to accurately assess AI-generated outputs \cite{stadler2024cognitive, lee2025impact}. These risks are particularly acute for medical students, for whom the development of robust reasoning and judgment is paramount. Furthermore, practical barriers such as patient privacy regulations and the substantial computational cost of local deployment often preclude the use of LLMs.

To counteract these risks, it is essential to develop frameworks that support exploratory learning and engagement with data, rather than passively accepting output \cite{safranek2023role, roustan2025clinicians}. Visual analytics, by integrating human expertise with machine intelligence in a “human-in-the-loop” paradigm, provides a promising foundation for such a framework. However, the application of visualization to this problem remains underdeveloped, much of the relevant work has focused on analyzing individual patient EMRs\cite{jin2020carepre, xu2023visual}, which does not facilitate higher-level organization of medical knowledge. Other studies have visualized structured medical text and other healthcare data, often neglecting the underlying semantics of the content\cite{palmer2020using, permana2024haiviz, siirtola2022glyph}, which also limits generalizability, as data standardization is often infeasible across different medical institutions.

To address these problems, we design MExplore (\cref{fig:mainInterface}), an entity-based interactive visual analytics approach for medical expertise acquisition. 
First, we extract MEs from unstructured medical text by leveraging automated text processing. These entities are strong indicators of text semantics\cite{gabrilovich2007computing, yamada2017learning, ling2020learning} and thus can represent the core knowledge within a text. Constructing a visual structure based on these entities can significantly accelerate the user's understanding of large-scale texts.
We then developed a visual analytic framework, integrating clutter and aggregating information of different densities based on medical paragraphs (MPs), medical entity sets (MESs), and MEs, enabling medical learners to explore at multiple levels, construct illness scripts, and acquire expertise. The contributions of this article include the following:

\begin{itemize}
	\item We propose a workflow that automatically extracts MEs from unstructured medical text, organizes them into a coherent structure, and generates semantic visualizations. It allows users to gain an overview of the knowledge structure and efficiently identify areas of interest.
	
	\item An innovative multi-level metaphor visual analytics framework is introduced. It enables users to explore knowledge at varying levels of granularity with low learning costs and guides them through the process of forming learning paths.
	
	\item Through case studies, user studies, and expert interviews, we demonstrate the system's effectiveness in analyzing medical texts and facilitating the acquisition of specialized knowledge.
\end{itemize}

Our design study involved an iterative process of design, evaluation, and collaboration among medical experts and visual analytics researchers. \hyperref[sec:related]{Section 2} reviews the relevant literature and prior research in this domain. \hyperref[sec:requirement]{Section 3} outlines the requirements and system workflow. \hyperref[sec:data]{Section 4} describes the data and processing. \hyperref[sec:mexplore]{Section 5} presents visual analytics design considerations and interactions. \hyperref[sec:evaluation]{Section 6} details the evaluation, including a case study and a user study. \hyperref[sec:discussion]{Section 7} presents the lessons learned from the study, outlines the limitations, and explores the potential generalizability of our approach. Finally, \hyperref[sec:conclusion]{Section 8} concludes the paper, summarizing the key findings and implications.

\section{Related work}
\label{sec:related}

\subsection{Medical expertise acquisition}
Recent research on medical expertise acquisition can be broadly categorized into two main areas: the construction of medical knowledge bases for knowledge structuring and the development of learning tools for illness script construction.

Efforts to build knowledge bases typically involve the integration and hierarchical processing of clinical documents \cite{lelong2019building, soualmia2011extracting} or medical literature \cite{kinast2023functional}. Some research has focused on disease-specific knowledge bases, designed to offer comprehensive coverage of particular diseases \cite{gong2019building}. These knowledge bases often include search functionalities, enabling users to retrieve relevant information or identify associations between terms \cite{landrum2020clinvar, soualmia2004combining}.

Various online tools have been developed to aid medical learners in forming illness scripts, including interactive reasoning systems\cite{zagury2022student}, gamified platforms, and educational videos\cite{hayward2016script, sophark2023clinical}. Other studies have focused on diagnostic training or case studies to support illness script development \cite{fall2021thinking, cloude2021designing}. More recently, tools such as ChatGPT have been explored for generating illness scripts in clinical settings \cite{yanagita2023assessing}.

In contrast to these studies, our method automates the extraction of MEs from unstructured medical texts, significantly enhancing efficiency compared with manual curation. Moreover, it incorporates rich data interactions and visual analytics, facilitating intuitive exploration and learning. Our approach illustrates the reasoning processes and offers more advanced interaction than one-sided output teaching or dialogue-based learning tools.

\subsection{Named entity recognition}

Named entity recognition (NER) is a task that involves identifying and classifying named entities in a text into predefined categories\cite{grishman1996message}. Extracting domain-specific concepts in specialized fields such as biomedicine is a specialized category of NER \cite{mehmood2023use}. The pedagogical value of such concepts in enhancing subject learning is well-documented\cite{nesbit2006learning}. Early NER approaches were predominantly rule-based, relying on manually curated lexicons and heuristic rules\cite{grishman1996message, hanisch2005prominer}. While effective in constrained settings, these methods lack scalability and struggle with domain adaptation. The emergence of deep learning transformed NER by enabling automated feature extraction, significantly improving recognition accuracy\cite{huang2015bidirectional, lample2016neural}. In particular, transformer-based architectures, such as BERT\cite{devlin2019bert}, further advanced the field by leveraging self-attention mechanisms to model complex semantic dependencies across text\cite{vaswani2017attention}. Large language models (LLMs) have recently introduced a generative paradigm to NER\cite{wang2023gpt}. However, despite their broader linguistic capabilities, LLM-based methods often underperform compared to transformer-based approaches\cite{han2023empirical}. This discrepancy may stem from the limited task-specific learning capacity of LLMs and the challenges associated with fine-tuning their extensive parameter spaces for optimal NER performance\cite{keraghel2024recent}.

While most efforts so far have focused on improving the accuracy of NER methods, relatively little attention has been given to their integration into expertise acquisition. Our approach addresses this gap by incorporating human-in-the-loop strategies and introducing a multi-layered, interactive analytical framework. Rather than merely presenting NER results as raw data or text, our method facilitates structured exploration, enabling users to understand better and engage with extracted entities.

\subsection{Visual analytics of medical text data}

Medical text data visual analytics research can be broadly categorized into medical literature data and EMR data\cite{wu2019evaluating, wanderer2016clinical}.

In medical literature, recent studies have focused on visualizing key terms. Approaches include the use of emerging named entities for knowledge discovery\cite{nawroth2020emerging}, and creating MeSH networks to analyze correlations between medical subject headings\cite{yang2018research}. Some studies have aggregated information from articles for further interaction\cite{mancini2022development}, explored disease theme development\cite{chen2020mapping}, and illustrating temporal changes in specific research areas, such as HIV\cite{dancy2018trends}. Additionally, medical knowledge maps have been developed to quickly locate and comprehend disease-related information\cite{zhou2022construction}.

Various visual analytics tools are developed to extracting actionable insights from EMRs. For instance, tools like ClinicalPath\cite{linhares2022clinicalpath} and CarePre\cite{jin2020carepre}, enhance clinical decision-making by visualizing test and treatment outcomes. DPvis\cite{kwon2020dpvis}, ChartWalk\cite{sultanum2022chartwalk} and other systems\cite{kenei2022supporting} aim to improve the efficiency of EMR review through the visualization of key clinical indicators. Furthermore, systems such as VIEWER\cite{msosa2023viewer} and DHIs\cite{jung2025designing} analyze user behavioral patterns by integrating visual representations of both textual and numerical data.

Compared with previous studies, our proposed method leverages real-world clinical text, focusing on the summarization and analysis of unstructured data from the perspective of medical entities. This enables the comprehensive utilization of widely available unstructured medical text, thereby enhancing generalizability. Moreover, we introduce a novel multi-level visualization analysis method that facilitates a hierarchical exploration process. Unlike single-level (e.g. the patient level) analysis, this approach enhances knowledge acquisition by enabling users to engage with data at varying levels of granularity, promoting efficient and intuitive exploration of medical text.

In summary, our approach addresses the limitations of previous studies by integrating ME into a visual analytical framework, enabling users to progressively build structured learning paths and acquire expertise.

\section{Requirement and system workﬂow}
\label{sec:requirement}

 \subsection{Requirements analysis}
We collaborated closely with ten domain experts (E1-E5). E1-E2 are PhD candidates specializing in medical studies, with 4 and 5 years of focused research experience; E3 is a professor at a medical college with 8 years of teaching experience; E4 is a medical researcher with a decade of research expertise; E5 is a dentist with 10 years of professional practice; and E6-E8 are experienced physicians with 5-10 years of clinical practice; E9-E10 are visualization researchers with 5-10 years of experience in visual analytics. 

E3-E8 contend that the acquisition of medical expertise involves the iterative refinement of knowledge into narrative structures, referred to as ``illness scripts''\cite{schmidt1993acquiring}. These scripts serve as psychological frameworks that organize clinical knowledge to facilitate efficient recall and application\cite{feltovich1984issues}. This framework enables clinicians to integrate relevant information, recognize patterns, and distinguish between different disease states\cite{custers2015thirty}. To support the development of these critical cognitive structures, through multiple rounds of iterative discussions with experts, we identified key analysis requirements as follows:

 \textbf{R1. Extraction of Core Knowledge Units from Complex Texts.}
 Unstructured medical texts are inherently dense and complex, presenting significant challenges for novice medical learners\cite{bagheri2023natural}. Due to their limited working memory capacity and developing cognitive schemas, identifying salient information from such texts imposes substantial extraneous cognitive load\cite{mancinetti2019cognitive}. This often results in a "reading bottleneck"\cite{perfetti1995cognitive}, where cognitive resources allocated to information identification and decoding detract from deeper comprehension and the effective transfer of knowledge into long-term memory. To mitigate this issue, it is essential to extract core knowledge units from extensive complex texts and eliminate redundant information, thereby reducing the additional cognitive load associated with navigating intricate material\cite{sweller1988cognitive, OEI2022cognitive}.
 
\textbf{R2. Organization of text with varying knowledge density to support gradual exploration.} 
For novice learners, information should be presented at appropriate levels of abstraction\cite{ganascia2015abstraction} and organized systematically to prevent the simultaneous presentation of excessive detail, which can lead to cognitive overload\cite{sweller1988cognitive, natesan2020clinical}. Therefore, medical text should be restructured with varying degrees of knowledge density and presented in a clear framework\cite{qiao2014using}, facilitating an incremental and exploratory learning process to comprehend complex medical expertise.

\textbf{R3. Revealing the Interconnections Between Knowledge Units.}
 In the learning process, rote memorization of facts is insufficient for deep comprehension and retention\cite{lujan2025paradox}. Learners must actively process new information, integrate it with existing knowledge structures\cite{mayer2025collaborative}, and organize it into coherent schemas to achieve long-term memory. To facilitate this crucial cognitive process, it is necessary to reveal the interconnections between knowledge units and deepen the understanding of their relationships.

  \textbf{R4. Focused Analysis of Key Knowledge Unit.}
The explicit identification of key knowledge unit is paramount for effective knowledge acquisition. By clearly delineating these essential concepts, learners can concentrate their analytical efforts and systematically organize their understanding\cite{etukakpan2025core}. These identified entities serve as cognitive "anchors", providing focal points around which learners can explore and deepen their comprehension of critical knowledge units\cite{medwell2020concept}. This directed attention not only streamlines the learning process but also facilitates divergent learning, allowing learners to expand their understanding and establish broader connections within the medical domain.

 \subsection{Workflow}\label{sec:workflow}
\begin{figure}[tb]
  \centering 
  \includegraphics[width=\columnwidth]{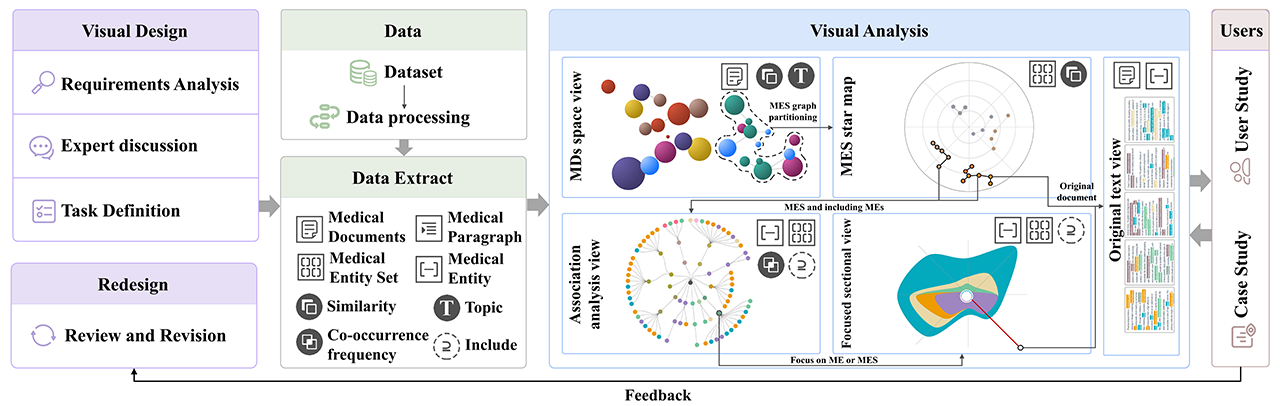}
  \caption{The MExplore workflow outlines a systematic process, including data extraction, relationship computation, visual analysis through a multi-level framework, and redesign through user feedback.}
  \label{fig:workflow}
\end{figure}

The workflow is depicted in Fig. \ref{fig:workflow}. Initially, the requirements were analyzed and discussed with domain experts to define the visual analytics tasks. The process begins by extracting ME and MES from the MD dataset through a structured data processing flow. We subsequently computed various relationships, including topic, similarity, co-occurrence, and containment. We propose a multi-level visual analytics framework enabling progressive exploration and expertise acquisition. The user first selects MDs within the MD space view. After segmentation and graph partitioning, the MP subgraphs are analyzed in the MP star map. Upon selecting the subgraphs, the association between the corresponding MESs and MEs is examined in the association analysis view. In this view, users can pick any node to focus on a sectional view for in-depth analysis. Throughout the analysis process, users can also add the original documents of the MP subgraph or MESs to the original text view for browsing and comparison. Finally, the system is redesigned and improved based on user and case study feedback.

\section{Data description and processing}\label{sec:data}
\subsection{Data description}
In this paper, we analyze and calculate two realistic datasets. The first dataset is the Chinese Biomedical Language Understanding Evaluation(CBLUE) provided by the Key laboratory of Computational Linguistics (Peking University, China) et al.\cite{zhang2021cblue}. CBLUE is collected from authorized medical textbooks and clinical practice, and contains 938 MDs and 47194 MPs. Each MP includes standard MEs, which are categorized into nine classes based on standardized chinese ME annotation\cite{zhang2021cblue}: disease (dis), clinical symptoms (sym), drugs (dru), medical equipment (equ), medical procedures (pro), body (bod), medical examination items (ite), microorganisms (mic), department (dep).

The second dataset is Medical text of West China School (Hospital) of Stomatology (MWCSS), collected between years 2023 and 2025, was provided by West China School (Hospital) of Stomatology (Sichuan University, China). This dataset contains around 100,000 MDs and 800,000 MPs which are all unstructured text data, including clinical diagnosis, treatment processes, specialist examicnations, auxiliary examinations, treatment plans, interventions, and drug instructions.

There is no personally identifiable information or offensive content involved in the texts of either dataset.

\subsection{Data processing}\label{sec:dataprocessing}

\begin{figure}[tb]
  \centering 
  \includegraphics[width=\columnwidth]{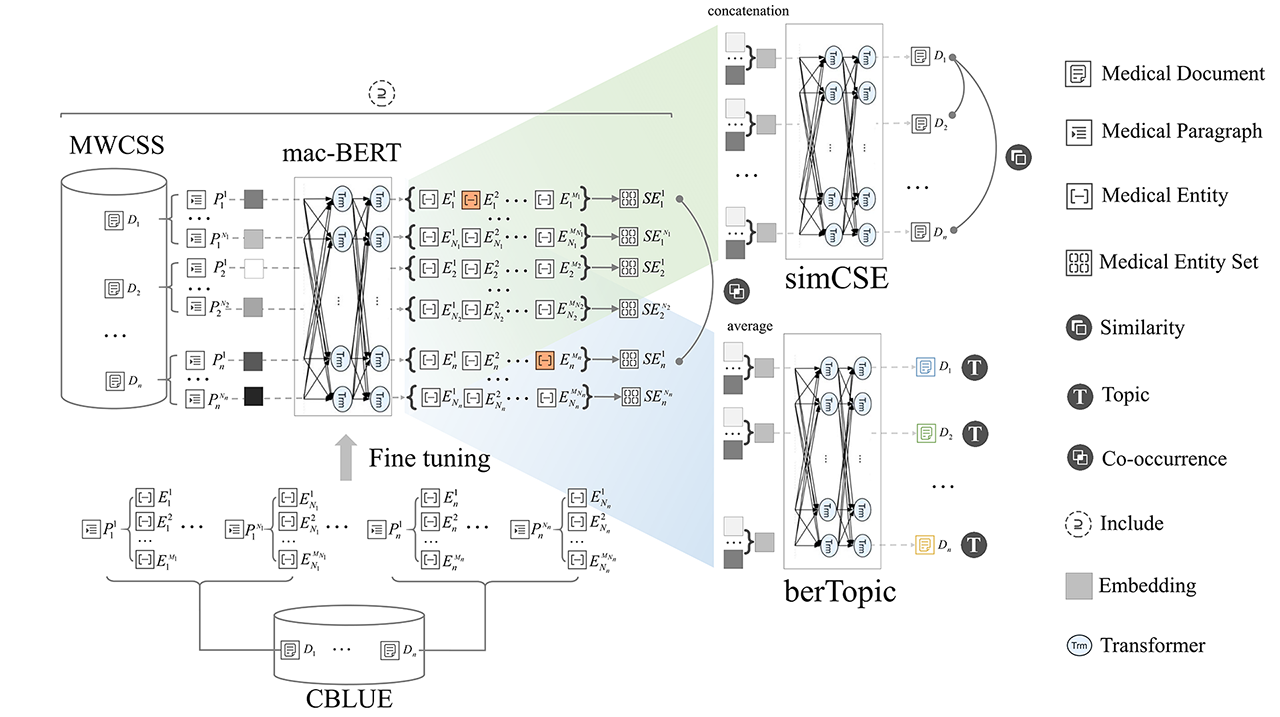}
  \caption{Pipeline of data processing. First, fine-tuning the Mac-BERT model is fine-tuned via texts in CBLUE. Each MP of the MDs is processed through the fine-tuned Mac-BERT model to extract MEs, which are subsequently organized into MESs subsequently. MES vectors are concatenated and input into SimCSE to assess document similarity. BERTopic is used for topic extraction by averaging MES embeddings to optimize topic word extraction, resulting in the topic classification of each document.}
  \label{fig:pipeline}
 \end{figure}

\ref{fig:pipeline} shows the pipeline of data processing

\textbf{1)} We employed MPs and MEs from the CBLUE dataset to fine-tune the Mac-BERT model, an improved BERT with novel masked language modeling (MLM) as a correction pretraining task\cite{cui2020revisiting}. Based on past benchmark tests, Mac-BERT performs excellently on the NER task, approaching human annotation levels\cite{zhang2021cblue}.

\textbf{2)} Each MD $D_i$ in both datasets contains $N_i$ MPs, and each MP $P^j_i$ is converted into a embedding vector and then put into the fine-tuned Mac-BERT model, to obtain MES $SE^j_i$. The relationship of the same entities in two MESs is called co-occurrence.

\textbf{3)} We perform concatenation on the embedding vectors of all $SE^j_i$ belonging to each document $D_i$ to merge them into a higher dimensional vector and input this vector into SimCSE (simple contrastive learning of sentence embeddings)\cite{gao2021simcse}, which is a framework for comparing text vector representations, obtained the similarity between documents.

\textbf{4)} Because BERTopic is more accurate than other methods\cite{grootendorst2022bertopic}, we adopt it for topic extraction and average the embedding vectors of all $SE^j_i$ in each document $D_i$ as the document embedding to optimize the topic word extraction further. Then we obtain the topic of each document.

\section{MExplore}\label{sec:mexplore}
\subsection{Visual analytics tasks}\label{sec:tasks}
Based on the analysis requirements (R1-R4), we identified the following visual analytics tasks (T1-T5). These tasks aim to help medical learners explore and acquire expertise through visual analysis.

\textbf{T1. Extract MEs from Real-World Medical Texts.} 
Extracting MEs as knowledge units from real-world medical texts enables learners to efficiently identify essential information. This process reduces extraneous cognitive load and facilitates the allocation of cognitive resources toward productive germane load, which is directly relevant to learning (R1). Furthermore, MEs can serve as foundational elements in constructing a coherent knowledge framework, supporting the development of knowledge structures.

\textbf{T2. Support for Cascading Visual Analysis and Exploration of Texts with Varying Knowledge Densities.}
Cascading visual analysis approach can organize and present medical entities (MEs) extracted at diverse knowledge densities, enables the exploration and analysis of medical texts across multiple levels of informational granularity, facilitating a seamless transition from high-level overview browsing to detailed, in-depth knowledge analysis as required. This capability can help learners facilitates the incremental exploration and understanding of complex medical knowledge (R2).

\textbf{T3. Establishing a Clear Structure for Association Analysis.}
To facilitate the integration of knowledge structures, it is essential to establish a clear and well-organized visual framework that explicitly maps the relationships between MEs and MESs. This structured representation enables learners to easily identify and analyze associations between various knowledge units (R3), thereby supporting schema formation\cite{mancinetti2019cognitive}. Through this process, learners can identify key knowledge units for further in-depth analysis(R4).

\textbf{T4. Support for entity-centric pattern analysis.}
To support the focused analysis of key knowledge units, the visual analytics framework must enable the analysis of contextual association patterns centered on MEs or MESs. ME-centered analysis allows learners to concentrate on core knowledge units (R4), fostering divergent exploration. This process not only deepens understanding but also facilitates the establishment of broader connections within the medical field\cite{medwell2020concept}, thereby reinforcing long-term retention.

\subsection{Visual design}\label{sec:desgin}

\begin{figure}[tb]
  \centering 
  \includegraphics[width=\columnwidth]{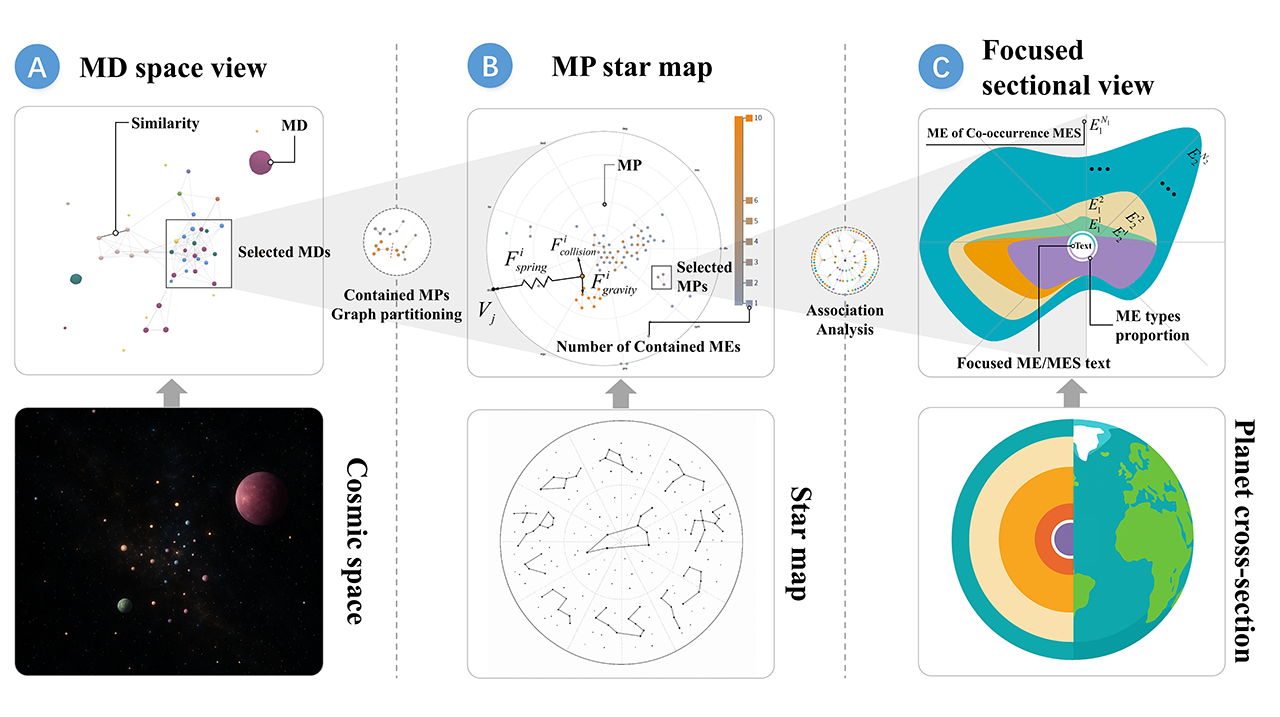}
  \caption{The MExplore framework: (A) the MD space view, inspired by cosmic space, where users select MDs and construct and partition graphs of the contained MESs; (B) the MP Star map, inspired by star map, where users choose corresponding MP subgraphs and perform relational analysis; (C) the focused sectional view, inspired by the planet cross-section, which enables detailed, focused analysis of the selected data.}
  \label{fig:framework}
\end{figure}

To address the requirements, tasks, and user-centered design principles\cite{munzner2009nested}, we design MExplore, a visual analysis system to assist medical learners in exploring and acquiring medical expertise. MExplore is based on a multi-layered metaphorical visual analysis framework, as illustrated in Fig. \ref{fig:framework}. This framework uses a three-level metaphor: cosmic space, star map, and planet cross-section, which inspires the design of corresponding visualization views: MD space view (Fig. \ref{fig:framework}A), MP star map (Fig. \ref{fig:framework}B), and focused sectional view(Fig. \ref{fig:framework}C). Each level supports visual analysis at different levels of granularity, allowing users to navigate progressively and intuitively through the knowledge domain (T2).

Rather than employing a zoomable interface, the system utilizes a coordinated layout that allows users to simultaneously compare and correlate information across different views while maintaining contextual coherence \cite{roberts2007state}. This layout enables users to switch between views, adjusting the analytical layer without losing sight of the overarching framework. By aligning the visual analytics process with intuitive metaphors rooted in common understanding, the framework facilitates the mapping of new information onto pre-existing cognitive schemas \cite{carroll1988interface}, thereby reducing cognitive load. Moreover, the multi-tiered, metaphorical structure, inspired by cosmic imagery, enhances memory encoding, fostering more effective knowledge retention as users interact with the system \cite{shneiderman2010designing}.

\subsubsection{MD space view}\label{sec:spaceview}
As shown in Fig.\ref{fig:mainInterface}A, after the keywords in the search bar are entered (Fig.\ref{fig:mainInterface}A\textsubscript{1}), the MDs containing the keywords are distributed in the MD space (Fig.\ref{fig:mainInterface}A\textsubscript{2}), where each MD is a celestial body whose size is proportional to the length of the text, and the color indicates the topic to which it belongs. The key terms of each topic are listed in Fig.\ref{fig:mainInterface}A\textsubscript{3} for quick identification of topic features. The 3D force-directed layout determines the positions of these celestial bodies, and the links between them indicate the gravitational force, which is proportional to their similarity. This metaphorical representation allows users to intuitively perceive complex relationships through spatial proximity and visual encoding, thereby reducing cognitive load \cite{sweller1994cognitive}. Compared with the 2D representation, the 3D design enhances structural comprehension by enabling depth navigation and providing a richer representation of knowledge density (T2). Users can further refine the visualization by adjusting a similarity threshold slider (Fig.\ref{fig:mainInterface}A\textsubscript{4}) to filter out weaker links. This interaction mechanism emphasizes core structural relationships, leveraging proximity to facilitate efficient identification of areas of interest \cite{ware2019information}. Selected MDs can then be explored in greater detail, supporting targeted and cognitively efficient knowledge discovery.

\subsubsection{MP star map}\label{sec:starmap}

Relying solely on the full-text embedding similarity between MDs to determine knowledge exploration pathways can introduce bias. For example, outpatient records, where certain sections—such as medical history and physical examination—tend to be more verbose than others, notably the diagnosis section. As a result, the embedding representation may disproportionately reflect longer sections, overshadowing concise yet critical information.

\begin{figure}[tb]
	\centering 
	\includegraphics[width=\columnwidth]{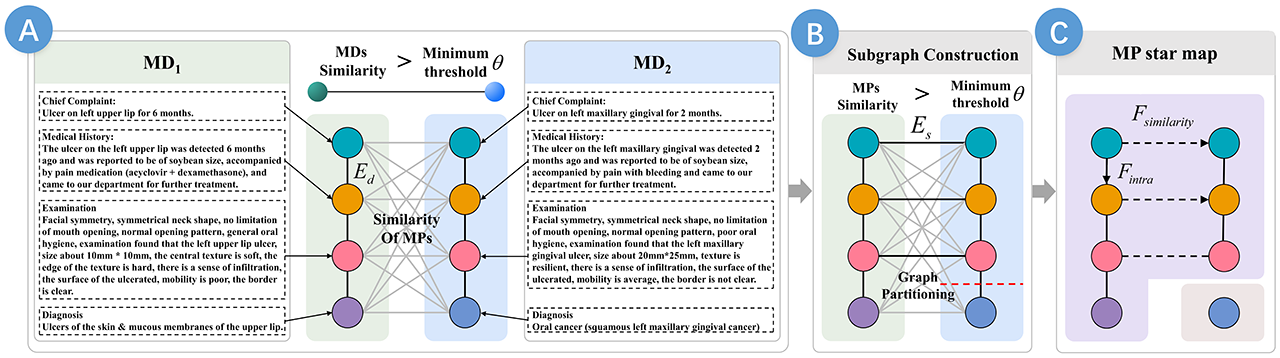}
	\caption{For detailed analysis, MDs are decomposed into MPs. (A) MPs are represented as vertices, with $E_d$ connecting MPs within the same MD, computing the similarity of MPs across different MDs. (B) MPs whose similarity is greater than $\theta$ are connected by $E_s$, and the graph is partitioned by KaFFPa. (C) The partitioned subgraphs are visualized via the constellation metaphor, where $E_d$s are shown as lines and MPs as stars. Each star is subject to $F_{\text{intra}}$ from $E_d$ and  $F_{\text{similarity}}$ from $E_s$.}
	\label{fig:graphPartitioning}
  \end{figure}

To address this limitation, we split MDs into MPs and model each MP as a graph vertex. MPs within the same MD are connected by intra-document edges $E_d$. We then computed pairwise similarity between MPs across linked MDs (Fig.\ref{fig:graphPartitioning}A). Pairs exceeding the similarity threshold set in Fig. \ref{fig:mainInterface}A\textsubscript{3} are considered semantically related and are connected by similarity edges $E_s$. The graph is subsequently partitioned via the KaFFPa algorithm \cite{sanders2011engineering}, which minimizes edge cuts and segregates MPs with distinct semantic content into separate subgraphs (Fig.\ref{fig:graphPartitioning}B).
Each resulting subgraph is visualized as a constellation in the MP star map (Fig. \ref{fig:framework}B), the MPs are depicted as stars, and $E_d$ is rendered as a constellation line, whereas $E_s$ is omitted to reduce clutter but is retained as latent factors influencing the layout, the similarity-based gravitational force $F_\text{similarity}$, and the structural gravitational force $F_\text{intra}$ from $E_d$ (Fig.\ref{fig:graphPartitioning}C). The combined gravitational force of star $i$ is calculated as follows:

\begin{equation}
	F_{\text{gravity}}^i = F_{\text{intra}}^i + F_{\text{similarity}}^i
\end{equation}
\begin{equation}
	F_{\text{similarity}}^i = \sum^{similarity^j_i > \theta}_{\text(i,j)} (\|\mathbf{F}\| \cdot similarity^j_i) \cdot \hat{\mathbf{d}}^j_i
\end{equation}
where $F_\text{intra}$ is the link force connecting the MPs within the same MD, with magnitude is a unit force $\|\mathbf{F}\| $. $\theta$ is the threshold set by the user, and $similarity^j_i$ is the similarity between MPs $i$ and $j$. $\hat{\mathbf{d}}^j_i$ is the unit vector of the force direction from MP $i$ to MP $j$.

In addition, each star $i$ is subjected to spring forces $F^i_\text{spring}$ from nine ME type poles $V_j$ uniformly distributed on the circular boundary of the star map:

\begin{equation}
	\mathbf{F}_{\text{spring}}^i = \sum_{j=1}^{9} \frac{num^j_i}{num_i} \cdot \|\mathbf{F}\| \cdot \hat{\mathbf{d}}^j_i
\end{equation}
where $num^j_i$ is the number of MEs of type $j$ in MP $i$, and $num_i$ is the total number of MEs in MP $i$. $\hat{\mathbf{d}}^j_i$ is the unit vector of force direction from MP $i$ to $V_j$.

To prevent occlusion, each star $i$ is subject to the collision force $F^i_\text{collision}$ of stars that may overlap it. The final combined force $F^i_\text{combined}$ is calculated as follows:
\begin{equation}
	F_{\text{combined}}^i = F_{\text{gravity}}^i + F_{\text{spring}}^i + F_{\text{collision}}^i
\end{equation}
This design was developed through an iterative process informed by feedback from an expert panel. Initially, we abstracted the entire subgraph into nodes to form a node-connected graph; however, this approach failed to convey the structure effectively. We then considered using a radviz plot but encountered issues such as occlusion. Finally, through this multi-force design, we achieved semantically coherent clustering ($F_\text{spring}$), representation of cross-document similarity and structural relationships ($F_\text{gravity}$), and clear visualization ($F_\text{collision}$) (Fig. \ref{fig:framework}B). To highlight containment for the same MD when the number of elements in the view is large, stars in the same constellation are assigned a color, whereas brightness gradients \cite{wang2008color} indicate the number of stars. Additionally, the border luminance of each star encodes the number of MEs contained within the MP, offering users a visual cue for assessing information density (T2).

In summary, this design offers a clear structural representation of MPs, aids learners in progressively constructing robust foundational schemas\cite{ganascia2015abstraction, qiao2014using}, thereby facilitating detailed, interactive exploration of the internal structure of MDs and promoting iterative, exploratory learning.

\subsubsection{Association analysis view}\label{sec:associationview}
Leveraging the concepts within MPs to create visual representations of hierarchical structures can significantly reduce the cognitive load and enhance the efficiency of knowledge navigation\cite{huang2019syllabus, zhou2024exevis}. Therefore, we extract MEs from each MP (T1), compose the MES, and visualize the resulting data via a radial tree structure. Compared with alternative text visualization methods, such as word clouds or Sankey diagrams, this approach effectively captures both inclusion and relational connections among elements. To enhance the visualization further, we propose an algorithm that groups MESs containing shared MEs into the same tree branch, thereby emphasizing co-occurrence relationships between the MESs (T3). The algorithm is outlined as follows:

\renewcommand{\algorithmicrequire}{\textbf{Input:}}
\renewcommand{\algorithmicensure}{\textbf{Output:}}
\begin{algorithm}
	\caption{Tree construction}\label{algo1}
	\begin{algorithmic}[1]
		\Require
		\begin{tabular}[t]{@{\hskip 9pt}p{10cm}l}
			1.Root node of the tree: $r_\text{tree}$ \\
			2.Set of nodes to be added: $N_\text{add}$
		\end{tabular}
		\Ensure 
		\begin{tabular}[t]{@{}l}
			Root node of the tree: $r_\text{tree}$
		\end{tabular}
		\For{each node $n_i$ in $N_\text{add}$}
			\State $C = \textit{GetMEs}(n_i)$
			\State $N, M_\text{checked} = \textit{TraverseTree}(r_\text{tree}, C)$
		\EndFor
		\State $N_\text{children} \gets N \cup M_\text{checked}$
		\For{each node $n_i$ in $N$}
			\State $\textit{RemoveFromTree}(r_\text{tree},n_i)$
		\EndFor
		\If{$L_N > 1$}
			\State $n_f \gets \textit{CommonFatherNode}(r_\text{tree},N)$
			\If{$n_f \text{ is not } (null \text{ or } r_\text{tree})$}
				\State $n_f.children \gets N_\text{children}$
			\Else
				\State $r_\text{tree}.children \gets$ $r_\text{tree}.children \cup N_\text{children}$
			\EndIf
		\EndIf
		\State \textbf{return} $r_\text{tree}$
	\end{algorithmic}
\end{algorithm}

The input consists of the root node of the tree, $r_\text{tree}$, and the set of nodes to be added, $N_\text{add}$. For each node $n_i$ in $N_\text{add}$, the algorithm retrieves its associated MEs using the $\textit{GetMEs}(n_i)$ function. It then traverses the tree through $\textit{TraverseTree}$ to identify and check for nodes intersecting with these MEs. During the traversal, nodes are checked recursively for intersections, and the set of relevant nodes, and the checked nodes are grouped into the set $N_\text{children}$ accordingly, which are subsequently added as children of the appropriate parent nodes.
If multiple nodes intersect, $N$, their common ancestor is identified by $\textit{CommonFatherNode}$, and the new nodes are added as children of this ancestor. The nodes are added directly to the root if no common ancestor is found. The process concludes by returning the updated tree structure.

We found that computational complexity of the algorithm is appropriate for the typical node count encountered during user study in \hyperref[sec:evaluation]{Section 6}. Given that the algorithm is primarily designed to construct association tree visualizations, an excessively large number of nodes could lead to confusion and increase the user cognitive load\cite{sweller1994cognitive}. Finally, we construct the association analysis view (Fig. \ref{fig:mainInterface}C), where each leaf node represents a ME, with the color indicating its type. Non-leaf nodes correspond to MESs, the colors of which are determined by a mixture of the type proportions of the MEs they encompass, thereby directly representing the type distribution within the MES. This view facilitates the perception of hierarchical and relational patterns among MEs and MESs, supporting schema construction and enhancing long-term memory retrieval\cite{dong2021some}. Furthermore, it assists learners in concept mapping, promoting clinical reasoning by externalizing and organizing cognitive processes.\cite{fonseca2024concept} Users can also add new medical paragraphs (MPs) at any time, enabling dynamic tree reconstruction and the seamless integration of new information into existing cognitive frameworks\cite{dong2021some}.

\subsubsection{Focused sectional view}\label{sec:focusview}

To delve deeper into a specific MES or ME for detailed analysis, we designed a focused sectional view (Fig. \ref{fig:framework}C). Inspired by the layered structure of a planet, the view positions the focused text as the core at the center. The middle donut chart encoding classification proportions forms the outer core, showing the proportion of MEs in different classifications, where $ \theta_{k} $ represents class $ k $. If the focus is on an ME, only one classification is displayed. The mantle layer visualizes co-occurring MESs through a polar-axis-aligned area chart. MES are divided into MEs, which are arranged based on their classification along the polar axis, where $ C_i^j $ represents the $ j_{th} $ ME in the $ i_{th} $ MES. The same classification of MEs of different MESs is connected by the area chart between the axes, where each axis's height is $ N_i^k * h_t $, which represents the number of MEs belonging to class $ k $ in the $ i_{th} $ MES. Compared to the design of an area chart, stream chart, etc., this view portrays the ME composition and rich semantic information of the MESs associated with the analyzed target. It generates distinct structural signatures that illustrate contextual association patterns of the focused ME/MES (T4). By employing dual encoding—simultaneously engaging imaginal and verbal systems—it enhances retention and recall through the creation of two complementary memory traces\cite{clark2023learning}. Furthermore, the mantle's radial patterns allow users to interactively compare MESs with similar ME configurations, fostering deeper cognitive engagement and more enduring impressions\cite{tversky2013visualizing}.

\subsubsection{Interaction}\label{sec:interaction}

\textbf{View exploration.} All the views in \hyperref[sec:desgin]{Section 4.2} support exploring interactions such as zooming and panning. The MD space view also allows users to rotate and drag, making it easier to explore large datasets within a limited space. By allowing users to directly interact with the data, MExplore transforms passive viewing into active exploration. The interactive nature of this design has been empirically shown to improve recall and long-term retention of intricate data structures \cite{barkley2020student}.

\textbf{Highlight and tooltips.} MExplore provides highlighting to make key data points stand out. Clicking on the data highlights them in the MD space and the MP star map. MEs in the original MD cards (Fig. \ref{fig:mainInterface}E) are colored for quick identification. The long text is shown in an abbreviated form to reduce visual clutter, with tooltips appearing on hover to display the full text for detailed analysis. This combination of highlighting and tooltips directs users' attention to essential information, supporting focused exploration without overwhelming them. This approach aids in cognitive load management, a critical factor for efficient learning and information retention\cite{sweller1994cognitive}.

\textbf{Data selection.} Users can filter and select MDs in the MD space view to create the MP star map. They can then click on specific subgraphs for association analysis. Next, users can click on the MES/ME node to focus on specific elements. Throughout the process, users can also select and view the original MD in Fig. \ref{fig:mainInterface}E for further comparison. This data filtering, selection, and comparison process mirrors decision-making and problem-solving activities typically found in active learning environments, can facilitating a deeper understanding of complex data relationships \cite{ibrahim2007impact}.

\section{Evaluation}\label{sec:evaluation}

In this section, we conducted three case studies with two domain experts (E3 and E5) and a user study with ten participants to demonstrate the usability and effectiveness of MExplore.

\subsection{Case study}\label{sec:casestudy}

Following an extensive discussion, the experts identified three key stages in the expertise acquisition process: discovery of areas of interest, association analysis, and construction of illness scripts. These stages were subsequently defined as the tasks for the case study. The study began with a 20-minute introduction to MExplore, after which participants were allowed to explore the system while concentrating on the tasks described above. The study lasted approximately one hour, and the experts were encouraged to ask questions or comment anytime.
Identification and exploration of areas of interest

\subsubsection{Identification and exploration of areas of interest}\label{sec:case1}

\begin{figure}[tb]
	\centering 
	\includegraphics[width=\columnwidth]{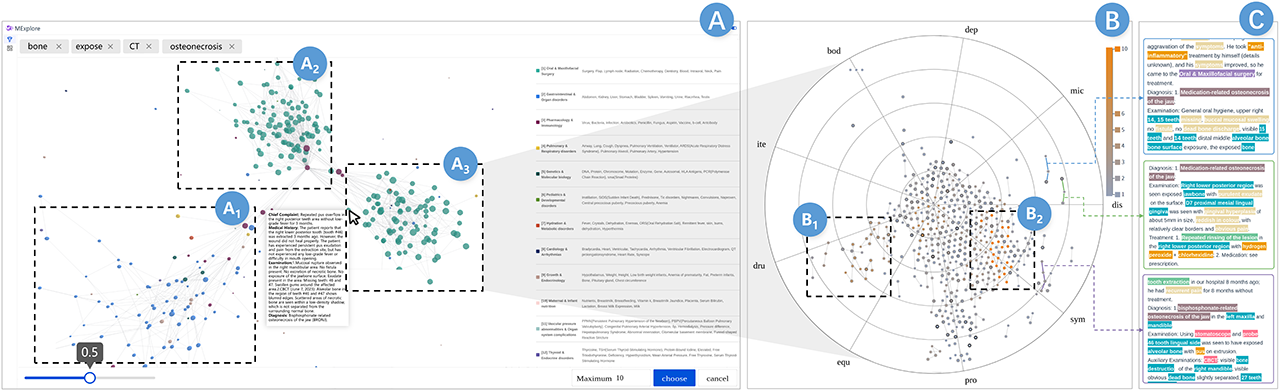}
	\caption{\textbf{A} Topics and distribution of keyword-related MDs, with a connection threshold set to 0.5. \textbf{B} Distribution of MP subgraphs within the MDs in A\textsubscript{2}. \textbf{C} Corresponding MD text for the subgraph of interest identified for further review.}
	\label{fig:casestudy_1}
  \end{figure}

The case study begins with a real patient with exposed bone in the facial area. A CT scan confirmed the presence of osteonecrosis. Experts entered the keywords \textit{bone}, \textit{expose}, \textit{CT}, and \textit{osteonecrosis} in MExplore. The related MDs create an MD Space. When the similarity threshold was set to 0.5, a clear structure appeared (Fig. \ref{fig:casestudy_1}A). There were three main star clusters: most MDs in  A\textsubscript{1} belonged to topic 2(\textit{Gastrointestinal \& Organ disorders}), while A\textsubscript{2} and A\textsubscript{3} were topic 1 (\textit{Oral \& Maxillofacial Surgery}). Because the exposed area was the face, the experts focused on topic 1.
A\textsubscript{3} contained MDs that described symptoms very similar to those of the patient. It was connected to A\textsubscript{1} and A\textsubscript{2} through the MDs of topic 3 (\textit{Pharmacology \& Immunology}). E3 selects all the MDs in A\textsubscript{3} and their connecting MDs in more detail. The maximum size of the subgraph was set to 10, which was found during their free exploration and works well for the segmentation of medical records. This result is shown in the MP star map (Fig. \ref{fig:casestudy_1}B). E5 first looks at subgraphs close to \textit{dis} and \textit{sym} to identify the specific condition. Three subgraphs were selected to review the text. After these MDs were compared with the original MDs (Fig. \ref{fig:casestudy_1}C), the experts concluded that the target disease was medication-related osteonecrosis of the jaw (MRONJ). E5 commented, "Compared with search engines, MExplore helps us find trustworthy MDs. Structured visual representations reduce cognitive overload and enable users to identify the target disease with interaction rather than search randomly through uncertain sources."

\subsubsection{Association analysis and identification of key factors}\label{sec:case2}

\begin{figure}[tb]
	\centering 
	\includegraphics[width=\columnwidth]{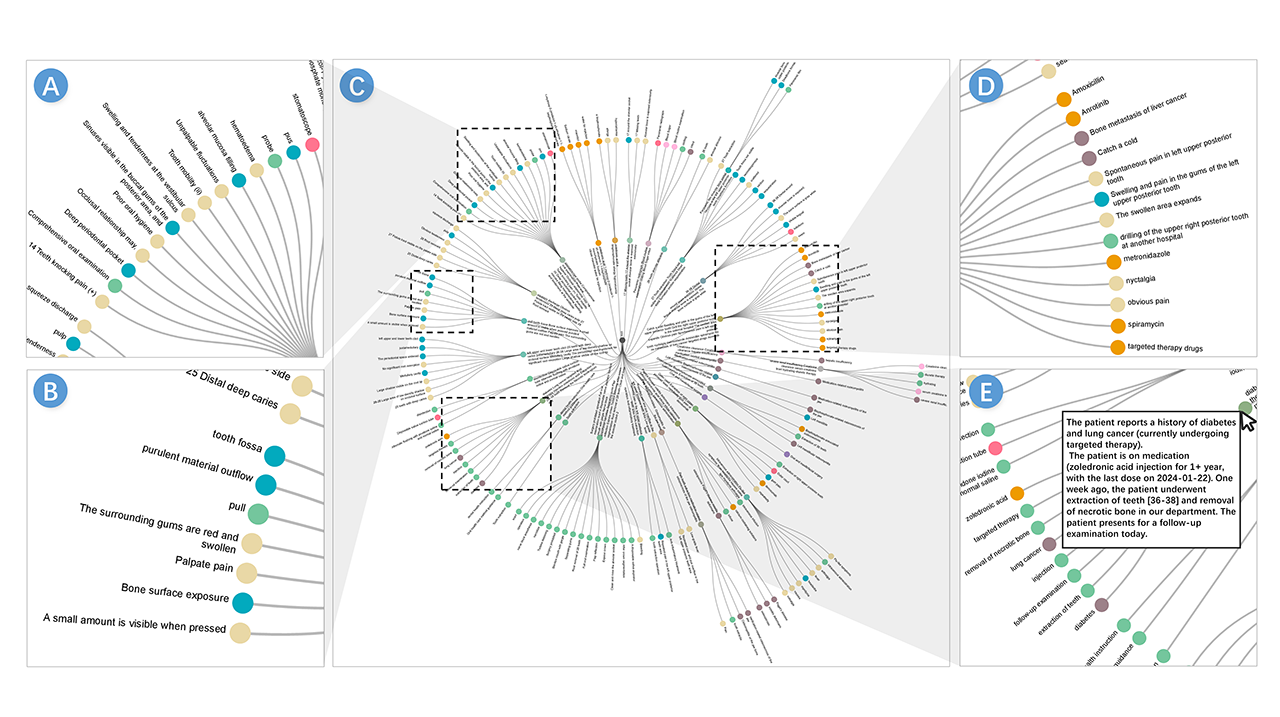}
	\caption{The association analysis view of the MES and MEs within the selected subgraph facilitates the exploratory analysis of relationships.}
	\label{fig:casestudy_2}
\end{figure}

Upon identifying the target disease, the experts conducted an association analysis to determine the key factors related to the illness script. As defined by Feltovich et al.\cite{feltovich1984issues}, illness scripts consist of three components: enabling conditions (EC), fault (FT), and consequences (CQ). In this case, the disease's name suggested a connection to the medication, leading the experts to select subgraphs that distribute bias toward \textit{dru} (Fig. \ref{fig:casestudy_1}B\textsubscript{1}). Additionally, the subgraphs adjacent to the target disease with a high total number of MEs were selected (Fig. \ref{fig:casestudy_1}B\textsubscript{2}). The selected subgraphs were organized into the association analysis view (Fig. \ref{fig:casestudy_2}C).

The experts reported that the MES in Fig. \ref{fig:casestudy_2}D had the highest number of \textit{dru} MEs, where \textit{targeted therapy drugs}, and \textit{Anrotinib} were not commonly used anti-inflammatory drugs, and \textit{Bone metastasis of liver cancer} indicated that the patients had cancer. In addition, Fig. \ref{fig:casestudy_2}E also mentioned \textit{lung cancer} and \textit{targeted therapy} and the patient used \textit{zoledronic acid}. Upon reviewing the corresponding MP, the experts noted that the patient had a history of cancer, had been undergoing long-term treatment, and had recently undergone a tooth extraction. Thus, E3 suggests that the cancer state and associated medication were factors related to EC and FT. For the CQ, E3 analyzed regions A and B in Fig. \ref{fig:casestudy_2} where \textit{bod} and \textit{sym} are more prevalent, and found that the main CQ factors were \textit{bone exposure} and \textit{exudate}. E3 concludes, "Performing disaggregation and association analyses on real-case factors can be integrated into clinical reasoning training, enhancing learners' clinical analysis skills".

\subsubsection{In-depth analysis and construction of illness scripts}\label{sec:case3}
\begin{figure}[tb]
	\centering 
	\includegraphics[width=\columnwidth]{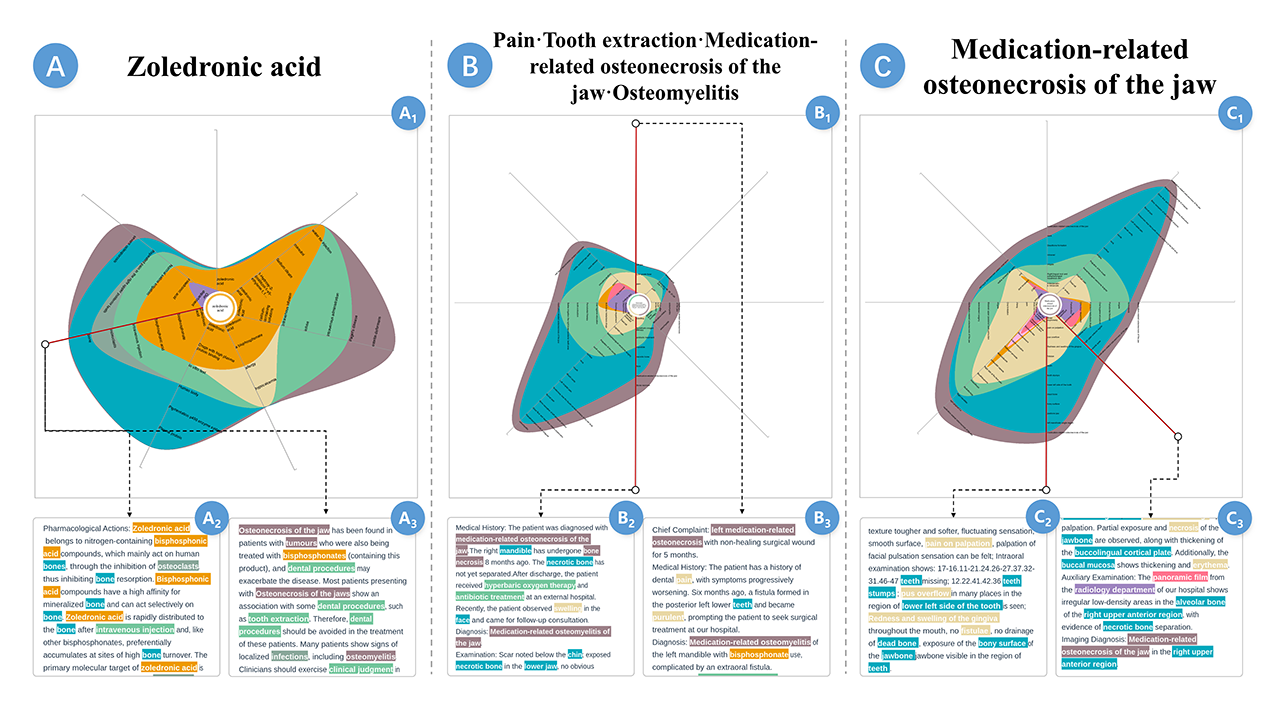}
	\caption{\textbf{A} Focused analysis of \textit{zoledronic acid} \textbf{B} Focused analysis of medical history MES including \textit{tooth extraction} \textbf{C} Focused analysis of \textit{MRNOJ}}
	\label{fig:casestudy_3}
\end{figure}

To construct a comprehensive illness script, the experts began by analyzing the factors presented in \hyperref[sec:case2]{Case 2}. E5 was initially interested in the role of \textit{zoledronic acid} as an EC for the development of MRONJ, so he focused on it and generated a focused sectional view (Fig. \ref{fig:casestudy_3}A\textsubscript{1}). An associated MES which includes \textit{osteoclasts} belonging to the \textit{mic} category, corresponding MD (Fig. \ref{fig:casestudy_3}A\textsubscript{2}), revealed that pharmacological action is to inhibit osteoclast growth, may impair bone repair. The MD also states that MRONJ may be caused when bisphosphonates (BPs) are used in patients receiving cancer treatment and is associated with dental procedures such as tooth extraction (Fig. \ref{fig:casestudy_3}A\textsubscript{3}). 
Subsequently, E3 analyzed the medical history (contains fewer \textit{pro} MEs), and MES containing \textit{tooth extraction}(Fig. \ref{fig:casestudy_3}B\textsubscript{1}). E3 observed that Fig. \ref{fig:casestudy_3}A\textsubscript{3} mentioned \textit{osteomyelitis}, and several co-occurring MES in Fig. \ref{fig:casestudy_3}B\textsubscript{1} also included it. A review of the original MDs (Fig. \ref{fig:casestudy_3}B\textsubscript{2} and B\textsubscript{3}) revealed that their chief complaint or medical history included MORNJ, suggesting that osteomyelitis may have developed as a CQ of MORNJ, potentially resulting from an infection.

To obtain a more definitive CQ, E5 performed a focus analysis of MRNOJ (Fig. \ref{fig:casestudy_3}C\textsubscript{1}), reviewing the MDs corresponding to the MES has \textit{sym} and \textit{equ} MEs (Fig. \ref{fig:casestudy_3}C\textsubscript{2} and Fig. \ref{fig:casestudy_3}C\textsubscript{3}). The analysis revealed that bone exposure and pus overflow were prominent clinical symptoms, while necrotic bone and low-density regions observed in the panoramic film emerged as distinctive imaging features. These findings can be utilized as reliable criteria for clinical diagnosis.
This analysis enabled the experts to construct a comprehensive illness script for MRONJ as follows:
\textbf{EC} Receiving treatment with BPs or other medications, undergoing cancer treatment, and having recently undergone dental procedures.
\textbf{FT} Medications such as BPs inhibit osteoclast function, while dental procedures, particularly tooth extractions, can cause trauma, both of which contribute to the onset of MRONJ.
\textbf{CQ} Exposed bone, pus drainage, and potential for subsequent infections, such as osteomyelitis.
As E3 noted that "MExplore enables learners to construct illness scripts in a short period, facilitating rapid diagnostic reasoning and enhancing long-term retention of script knowledge."

\subsection{User study}\label{sec:userstudy}
The purpose of this user study was to evaluate the effectiveness of the MExplore in facilitating the acquisition of medical expertise. The study design was informed by previous research\cite{keemink2018illness, Moghadami2021Teaching, Young2016Influence, XU2023}  that employed illness script construction tasks to assess medical expertise acquisition. We invited 20 undergraduate medical students in year 2 (10 male,10 female) to participate in the study. A randomized controlled trial compared both the accuracy and efficiency of illness scripting tasks via the MExplore with those of the traditional learning method. Additionally, a follow-up test was conducted two weeks later to assess the impact of MExplore on long-term expertise retention\cite{cepeda2006distributed}.

The experimental procedure proceeded as follows: We first introduced MExplore, detailing its purpose and features, followed by a training session on its usage. The participants were given 20 minutes to explore freely and understand the meaning of each view. Subsequently, they were tasked with constructing illness scripts.

For the illness script task, we selected three diseases—oral candidiasis (D1), meningitis (D2), and herpes zoster (D3)—that are relevant to our panel's area of expertise. These diseases were chosen from a list of diseases for which illness scripts could be activated\cite{custers1998role}. Importantly, the participants had not previously studied these diseases systematically. Following prior studies\cite{keemink2018illness}, the task began by providing the participants with a brief description of typical cases for each disease, including relevant medical history and examination results, excluding the disease name. Based on this information, participants were asked to identify the disease and complete an illness script template, the information of which was divided into three categories: EC, FT, and CQ\cite{custers1998role,custers2015thirty}. Each correct information unit was awarded one point. For instance, "this disease is common in middle-aged women and presents with unilateral headaches," the response would be scored as five points: two points for the EC (age and sex), and three points for the CQ (symptom (pain), organ affected (head), and the location of pain (unilateral)). Accuracy was calculated by comparing participants' responses to standardized answers agreed by the panel, with the best possible score set at 100\%\cite{bland2005psychometric}.

\begin{figure}[tb]
	\centering 
	\includegraphics[width=\columnwidth]{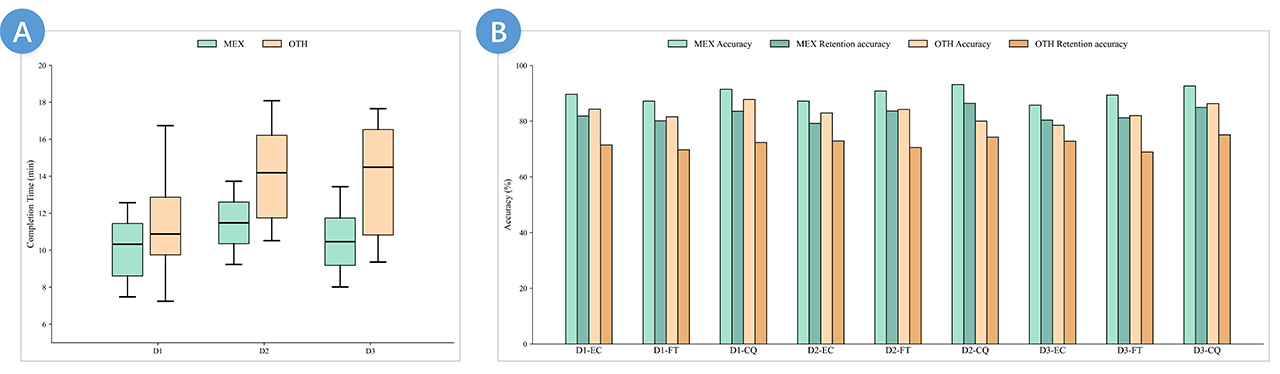}
	\caption{User study results. \textbf{A} Comparison of completion times. \textbf{B} Comparison of accuracy and retention accuracy. ME: Use MExplore for expertise acquisition; OTH: Acquiring expertise without MExplore but other resources.}
	\label{fig:userstudy}
\end{figure}

The participants were randomly assigned to one of two groups, each consisting of five males and five females. One group utilized the MExplore (MEX), while the other group was free to consult external resources, such as textbooks and case retrieval, without using the system (OTH). We assessed both groups based on the accuracy of their generated illness scripts and the time required to reach a conclusion. Additionally, participants were asked to refill their answer sheets two weeks later based solely on recall to record retention accuracy. The results of the study are presented in Table \ref{tab:user_study} and Fig. \ref{fig:userstudy}.

Table \ref{tab:user_study} presents the average accuracy, retention accuracy, and completion time for each disease across the different information classification methods under both experimental conditions. A more detailed comparison of these metrics is provided in Fig. \ref{fig:userstudy}, which includes box plots for completion time and bar charts for accuracy and retention accuracy.
The results show that MEX leads to better accuracy, retention, and faster task completion than OTH. For all datasets (D1, D2, D3), the MEX group consistently had higher accuracy, with differences ranging from 3–6\% higher than those of OTH. The ME group also outperformed OTH by 6–13\% in terms of retention accuracy, likely owing to the interactive process of exploring and reflecting on real-world cases. Additionally, the MEX group completed tasks faster, with an average time that was 1.5–3 minutes shorter than that of the OTH group, and their completion times are more stable (Fig. \ref{fig:userstudy}A). This efficiency can be attributed to the system's structured exploration path, which reduces the potential for time loss due to disorientation.

\begin{table}[tb]
	\caption{User study results.}
	\label{tab:user_study}
	\scriptsize%
	\centering%
	\begin{tabu} to \columnwidth {  
		X[1.5,c]  
		X[1,c] X[1,c]  
		X[1,c] X[1,c]  
		X[1,c] X[1,c]  
	  }
	  \toprule
	  \multirow{2}{*}{\shortstack{Information\\ Category}} & 
	  \multicolumn{2}{c}{Accuracy (\%)} & 
	  \multicolumn{2}{c}{\shortstack{Retention \\ accuracy (\%)}} & 
	  \multicolumn{2}{c}{\shortstack{Completion \\ time (min)}} \\
	  \cmidrule{2-3} \cmidrule{4-5} \cmidrule{6-7}
	  & MEX\footnotemark[1]
	  & OTH\footnotemark[2] & 
	  MEX & OTH & 
	  MEX & OTH \\  
	  \midrule
	  \textbf{D1} & & & & & & \\  
	   \hspace{1em}EC & 89.67 & 84.32 & 81.90 & 71.45 & \multirow{3}{*}{10.45} & \multirow{3}{*}{12.32} \\
	   \hspace{1em}FT & 87.21 & 81.56 & 80.15 & 69.75 & & \\  
	   \hspace{1em}CQ & 91.45 & 87.80 & 83.56 & 72.38 & & \\
	  \textbf{D2} & & & & & & \\  
	  \hspace{1em}EC & 87.23 & 82.98 & 79.21 & 72.90 & \multirow{3}{*}{13.40} & \multirow{3}{*}{13.85} \\
	  \hspace{1em}FT & 90.82 & 84.21 & 83.67 & 70.59 & & \\  
	  \hspace{1em}CQ & 93.15 & 80.02 & 86.43 & 74.28 & & \\
	   \textbf{D3} & & & & & & \\  
	   \hspace{1em}EC & 87.75 & 78.56 & 80.38 & 74.81 & \multirow{3}{*}{11.82} & \multirow{3}{*}{15.05} \\
	   \hspace{1em}FT & 89.40 & 82.01 & 81.21 & 68.96 & & \\  
	   \hspace{1em}CQ & 92.67 & 86.34 & 84.98 & 75.12 & & \\
	  \midrule
	\end{tabu}
  \end{table}
  \footnotetext[1]{Acquiring expertise using MExplore.}
  \footnotetext[2]{Acquiring expertise without MExplore but other resources.}

\subsection{Expert interviews}\label{sec:interview}
\begin{figure}[tb]
	\centering 
	\includegraphics[width=\columnwidth]{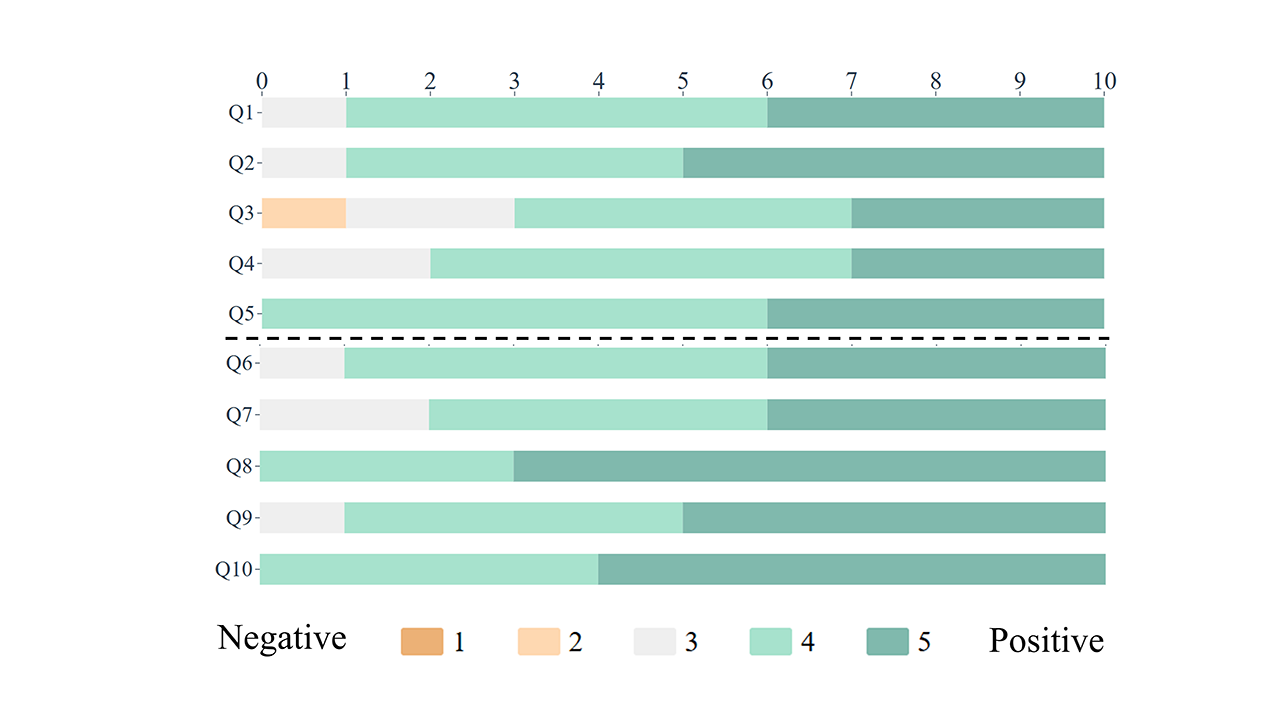}
	\caption{The association analysis view of the MES and MEs within the selected subgraph facilitates the exploratory analysis of relationships.}
	\label{fig:expertInterview}
\end{figure}

We conducted semi-structured interviews with domain experts in \hyperref[sec:requirement]{Section 3}, each experts completed a questionnaire with 11 items (see Table \ref{tab5}), using a five-point Likert scale to assess their attitudes toward the system\cite{likert1932technique}. The results of the questionnaire are presented in Fig. \ref{fig:expertInterview}, following the statistical method outlined by \cite{lin2021towards}.

\begin{table}
	\begin{center}
		\begin{minipage}{\columnwidth}
			\caption{User questionnaire}\label{tab5}
			\begin{tabular*}{\columnwidth}{c | p{22em}}
				\toprule%
				Q1 & MExplore is very easy (difficult) to learn. \\
				Q2 & MExplore is very easy (difficult) to use. \\
				Q3 & The visual design of the MExplore are easy (difficult) to understand. \\
				Q4 & The visual interactions of the MExplore are easy (difficult) to use. \\
				Q5 & I am very willing (unwilling) to use MExplore in exploring and acquiring medical expertise. \\
				\midrule
				Q6 & Using MExplore, I can (cannot) to efficiently identify core knowledge units within complex medical texts. (R1)\\
				Q7 & Using MExplore, I can (cannot) to explore medical texts in a gradual, structured manner. (R2)\\
				Q8 & Using MExplore, I can (cannot) to identify and analyze the interconnections between knowledge units. (R3)\\
				Q9 & Using MExplore, I can (cannot) to focus on and analyze key knowledge units in detail. (R4)\\
                Q10 & MExplore can (cannot) helps me consturct comprehensive and reliable illness script. \\
				\bottomrule
			\end{tabular*}
			\footnotetext{Q1-Q5 focus on assessing the system performance of MExplore, Q6-9 evaluate whether the key analysis requirements(R1-R4) are satisfied, and Q10 is the overall objectives.}
		\end{minipage}
	\end{center}
\end{table}

\textbf{System performance.} The experts all recommended MExplore for its user-friendly interface and interaction capabilities. The system's hierarchical visualization, organized by data granularity and incremental exploration ability, significantly enhances user understanding and analysis. Although some of the experts(E6-E8) had no prior experience with visual analytics systems, they could easily comprehend and effectively utilize the system after minimal training. E4 concurred that the ME-based exploration learning process in MExplore allows learners to focus on the core concepts of the target domain while creating an independent exploration path. E2 suggested incorporating features such as exploration playback and snapshot functionality to further increase the system's usability. In summary, the experts highly praised the system's visual design and interaction approach, asserting that it holds significant potential to improve the efficiency of knowledge acquisition for medical professionals.

\textbf{Analysis requirements.} 
As shown in Fig. \ref{fig:expertInterview}, experts believe that most of the key analytical requirements have been well met. E9 noted that the design of the association analysis view and the focused sectional view enabled users to correlate and concentrate their analysis on key knowledge points, thereby identifying relevant and valuable information—an essential aspect of the knowledge acquisition process. E8 noted that compared with search-based learning methods or the chatbots that have become popular in recent years, this approach can exercise the learner's thinking ability and improves knowledge retention and accumulation. E5 further highlighted that "In addition to assisting novices in acquiring expertise, MExplore can also support experts and researchers in disease studies by aiding in research, summarization, and the discovery of new features."

\section{Discussion}\label{sec:discussion}

\textbf{Lessons learned.} The key insight from this study is the importance of comprehensibility. Learners must be able to focus on acquiring knowledge throughout the learning process, and the visualizations and interactions should be designed in a way that does not overwhelm or distract users. Through an iterative design process involving both experts and users, we find a clear preference for simple and intuitive visual forms. Based on this feedback, we developed a multi-level metaphor visual analytics framework to reduce cognitive strain and facilitate a paradigm shift from traditional learning methods to more interactive forms of knowledge acquisition.

\textbf{Limitation.} One limitation of this study is the limited scope of the dataset. While Mexplore can be applied to datasets of different diseases in a short time, integrating additional datasets will further enhance the system's capabilities. Moving forward, we plan to expand the dataset through ongoing data accumulation and collaboration with additional organizations, thereby broadening the range of diseases covered and enhancing the applicability of Mexplore. Furthermore, some users have reported latency when processing large-scale MDs and MPs. To address this, we intend to iterate on the system by optimizing algorithms and testing with larger datasets to reduce lag and improve overall user satisfaction.

\textbf{Generalizability.} Although MExplore is designed for medical expertise acquisition, its methodology can be extended to other domains requiring structured knowledge exploration. For example, annotated legal texts could be used to fine-tune predictive models, extracting relevant entities and structures for legal expertise acquisition. These structured data sources could then be integrated into MExplore for visual analysis and exploration, facilitating a similar expertise acquisition process in legal research and other disciplines.

\section{Conclusion}\label{sec:conclusion}
In this paper, we introduced MExplore, a visual analysis tool designed to support medical learners in exploring and acquiring medical knowledge. MExplore extracts MEs from large-scale medical texts and constructs a knowledge structure, offering a multi-level metaphor visual analytics framework. This framework includes four coordinated views: the MD space view, the MP star map, the association analysis view, and the focused sectional view. Together, these views allow users to customize their learning paths, promoting deeper comprehension, and enhanced retention. By facilitating the construction of complex, interconnected schemas typical of medicine, MExplore significantly supports learners in acquiring medical expertise.

We conducted three case studies, a user study with real-world medical data, and expert interviews. The results highlight the effectiveness of MExplore in facilitating the exploration and analysis of medical texts, significantly enhancing the learning experience and supporting the acquisition of medical expertise.


\bibliographystyle{abbrv-doi}

\bibliography{MExplore}
\end{document}